\documentclass[a4paper,fleqn,usenatbib]{mnras}

\usepackage[T1]{fontenc}
\usepackage{ae,aecompl}

\usepackage{graphicx}	
\usepackage{amsmath}	
\usepackage{amssymb}	
\usepackage{enumerate}

\title[Analysis of the Kinematic Structure of the Cygnus OB1 association]{Analysis of the Kinematic Structure of the Cygnus OB1 association}

\author[M. T. Costado et al.]{
M. T. Costado$^1$,\thanks{E-mail: mteresa@iaa.es}
E. J. Alfaro$^1$,
M. Gonz\'alez$^1$
and L. Sampedro$^{1,2}$
\\
$^1$Instituto de Astrof\'\i sica de Andaluc\'\i a, CSIC, Apdo 3004, 18080 Granada, Spain\\
$^2$Instituto de Astronomia, Geof\'isica e Ci\^encias Atmosf\'ericas, Universidade de S\~ao Paulo, 05508-090, S\~ao Paulo, Brazil.  
}
\date{Accepted 2016 November 11. Received 2016 November 11; in original form 2016 September 23}

\pubyear{2016}

\begin{document}
\label{firstpage}
\pagerange{\pageref{firstpage}--\pageref{lastpage}}
\maketitle

\begin{abstract}
The main objective of this study is the characterization of the velocity field in the Cygnus OB1 association using the radial velocity data currently available in the literature. This association is part of a larger star-forming complex located in the direction of the Cygnus region, but whose main subsystems may be distributed at different distances from the sun. We have collected radial velocity data for more than 300 stars in the area of 5 x 5 square degrees centred on the Cygnus OB1 association. 
We present the results of a kinematic clustering analysis in the subspace of the phase space formed by angular coordinates and radial velocity using two independent methodologies. 
We have found evidence of structure in the phase space with the detection of two main groups, corresponding to different radial velocity and distance values, belonging to the association, and associated with two main shells defined by the $H{\alpha}$ emission. A third grouping well separated from the other two in velocity appears to occupy  the whole region associated with what has been called "common shell".

\end{abstract}




\section{Introduction}

Looking at the Cygnus constellation, it appears that we are looking down a spiral arm, with gas and young stars distributed at several distances (from a few hundred parsecs to 1 - 2 kpc and even well beyond) along the line of sight. However, establishing the number and location of Galactic spiral arms is far from being free of controversy. The Cygnus region is a clear example of this situation. Looking at the Galactic Longitude $\approx$ 80$^o$, the line of sight could be either tangent to the Local {\it armlet}, or secant to two major arms, depending on the Galactic model adopted \citep{VA2011, VA2014}. What is clear is that the Cygnus constellation harbours one of the most interesting star-forming complexes, which, together with the Carina region, displays the largest surface density of massive stars. In both cases, we could be looking along a tangent line of a spiral arm.

A proof of this has been reported by \cite{ST2014} analysing the distance distribution of the stellar population in the M29 cluster area, where the stars are in the field range of between a few hundred parsecs up to more than 2 kpc. M29 (NGC 6913) forms part of the Cygnus OB1 association, which also contains five other young clusters. This OB association presents a complicated 3D geometry, and of the nine associations forming the Cygnus-X region is the one with the largest distance interval along the line of sight.  This is our target object, the Cygnus OB1 association, and we will try to analyse the distribution of the stellar population in a subspace of the phase space formed by angular coordinates (RA, DEC) and radial velocity (RV), studying how many stellar populations with different kinematic behavior can be found in the region, and how these stellar groupings, if any, might be associated with the third spatial coordinate: the distance to the Sun.

There are some previous studies about kinematic structure in the RV subspace of stellar systems, for example, for clusters as NGC 2264 \citep{FU2006}, Orion Nebula Cluster \citep{FU2008}, Gamma Velorum \citep{JE2014} and NGC 2547 \citep{SA2015}; and for associations as Puppis OB \citep{PE1981}, Perseus OB2 \citep{BE2002} and Cygnus OB2 \citep{WR2016}. But these kinds of studies are few due to the lack of precise and complete kinematic data for the clusters as well as the absence of automatic tools specifically designed for the search for kinematic structures. 
A good characterization of the phase space is becoming the most reliable tool for detecting and isolating different stellar  populations in star systems. 
We have recently developed an automatic tool for performing a clustering analysis of the phase space \citep{AL2016}. 
In addition, we will make use of another cluster analysis tool called OPTICS \citep{ES1996, AN1999} to look for different phase-space structures in the Cygnus OB1 association.

The paper is therefore structured as follows. In section 2, we describe some previous studies related to the Cygnus OB1 association, what kind of data we are using and how they were selected, and the stellar objects in the studied area. The kinematic structure of this association in the phase subspace is presented in section 3, while a description of the kinematic populations of this star-forming region is discussed in section 4. Lastly, the main conclusions of this work are summarized in section 5.

\section{The Cygnus OB1 association: Antecedents, data and stellar populations. }

\subsection{Background}

An OB association is a gravitationally unbound stellar system, containing between 10 and 100 massive stars of early spectral types (O and B stars) and possibly hundreds or thousands of low- and intermediate-mass stars. Associations are believed to form within a relatively small region of space inside a giant molecular cloud. After the removal of gas and dust, the remaining stars become unbound and, in most cases, begin to expand. The typical age of an OB association is 10 Myr,
and the typical size is around 80 pc \citep{EL2009}. It is believed that the majority of all stars in the Milky Way were formed in OB associations. 
A census of the stellar content of the OB associations within 1 kpc from the Sun was presented by \cite{ZE1999}, based on {\it Hipparcos} positions, proper motions and parallaxes. 

\begin{figure}
\centering
\includegraphics[scale=0.75]{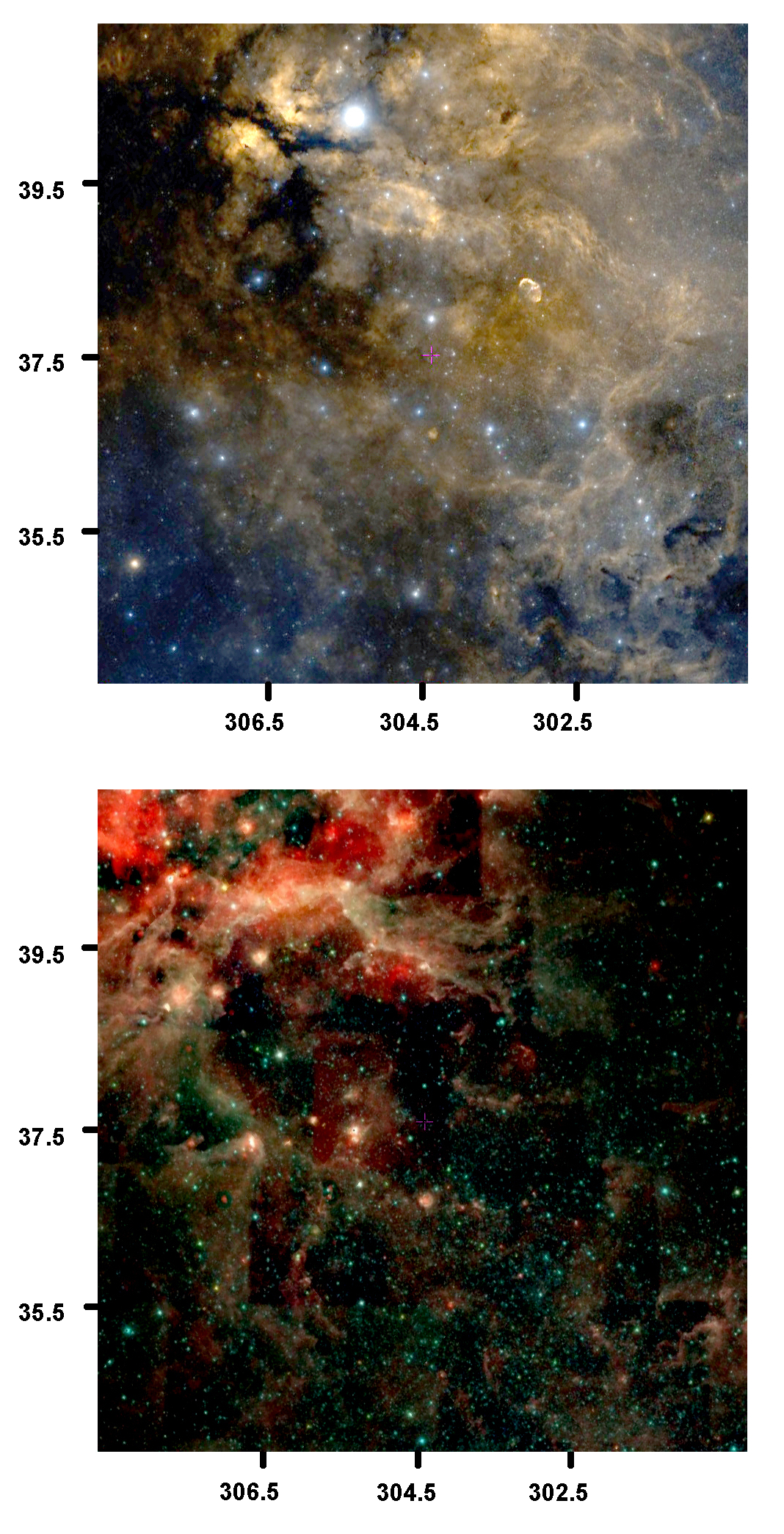}
\caption{Digitized Sky Survey coloured image ({\it top}) and ALLWISE colour image ({\it bottom}) centred on the Cygnus OB1 association (RA = 304.45 deg, DEC = +37.63 deg) taken from the Aladin server within a square 7.5 degrees size each. North is up and East is left. }
\label{fig:location}
\end{figure}

The Cygnus-X complex is approximately 10 deg in diameter and  is located at an average distance of around 1 kpc, containing several active star-forming regions. There are nine OB associations listed in the Cygnus region, seven of which are likely physically connected to the super bubble centred on the Cygnus OB2 association (Galactic longitude $\approx 80^o$). We are interested in studying the Cygnus OB1 association, specifically we want to analyse the kinematic structure of the association, which is located at the galactic coordinates ($l \sim 75.5^o$, $b \sim 1.1^o$). Fig, \ref{fig:location} shows two images of the association, in optical ({\it top}) and infrared ({\it bottom}) taken from the Aladin server\footnote{http://aladin.u-strasbg.fr/}, where they show the complex structure of the stellar, gas, and dust distributions in the 7.5 degrees size selected area.

\cite{AR2013} analysed the interstellar medium (ISM) in the line of sight of the Cygnus OB1 association by studying a family of cometary globules (CGs) detected by {\it Spitzer} . The CGs family considered could be a remnant of a pillar directed towards the centre of the association. The pillar is decaying in the process of star formation under the action of ionizing radiation and the wind of the OB stars.

\cite{CH2002} concluded that a rich OB association can produce a set of expanding shells: a massive slow-expanding one swept up by the stellar winds of massive stars and several fast inner shells produced by the activity of Wolf-Rayet (WR) stars and, later, by supernova explosions. They argued that these dynamical structures provide favourable conditions for triggering star formation. 
Lozinskaya and collaborators \citep{LO1988, LO1998, LO1999, CH2002} studied the effects of stellar winds on the interstellar medium using H$\alpha$ line plus IRAS and/or CO data. The super shell around the OB associations has the coexistence of bright emission at low velocities with weak high-velocity emission. The densest clouds of the Cygnus OB1 association are at a distance of 1.2 to 1.4 kpc and the density of O - B2 stars has a maximum at the same distance, showing the relation between the dense cloud and the stars. The high-velocity gas in the slow super shells is accelerated by shock waves generated by the winds from WR stars and by supernovae that interact with the tenuous gas of the inner cavity, which induces a new wave of star formation in the parent molecular cloud. You can see the hierarchical system of shells for the Cygnus OB1 association in figure 1 of \cite{LO1998}. This figure reveals that the northeastern part of the common shell around the cluster NGC 6913 is greatly attenuated in optical emission and that this dense cloud coincides with the dust lane at a distance of 1 - 1.2 kpc. Moreover, in the southwest part of the common shell around the cluster IC 4996, there is a deep cavity. These two boundary regions of the common shell overlap with some Cygnus OB9 and OB3 stars. Besides that, there are some small-scale shells around individual WR and Of stars. 

\cite{SC2007} carried out a multi-wavelength study about the Cygnus complex, in which they found evidence that three molecular clouds in that region are directly shaped by the UV radiation mainly from members of NGC 6913. There is a cavity around this object, which was created by radiation and/or stellar wind compression from massive stars in the region. 

The super shell surrounding the Cygnus OB1 association was studied in the far-infrared wavelength using IRAS data by \cite{SA1992}. They found that the association is well defined covering an area of 2 x 5 degrees and deficient in IR emission with a limb edge and a size of 50 x 130 pc at 1.5 $\pm$ 0.3 kpc of distance. They said that this cavity represents the early stages of a super bubble produced by the winds and supernovae from spatially distributed massive stars. 

 \cite{SI2015} studied the periphery of the super shell surrounding the Cygnus OB1 association and, specifically, the stellar cluster vdBergh 130 and the ISM around it. They concluded that this cluster is physically connected with the super shell around the association, and that there are some ionized regions of bright mid-IR emission in the vicinity of the object. They only estimated the minimum colour excess as their study showed a large scatter of colours, as well as the cluster age range of 5 - 10 Myr and distance of 1.8 $\pm$ 0.3 kpc. Taking into account all the previous studies, we assume a distance between 1 and 1.8 kpc for the Cygnus OB1 association in the following.

\subsection{Radial velocity measurements}
\label{sec:file} 

We first did a basic search of RV data from the VIZIER service\footnote{http://vizier.u-strasbg.fr/viz-bin/VizieR} centred on the Cygnus OB1 association (RA = 304.45 deg, DEC = +37.63 deg) using a box with a side of 5 degrees. We have chosen this big area to be sure that the Cygnus OB1 association is completely covered, although some contamination from the Cygnus OB3 and OB9 associations will be present, since the borders between them are not clear. We found five independent catalogues: \cite{BA2000}, \cite{AB1972}, \cite{WI1953}, \cite{FE1996} and \cite{EV1967}. 
We downloaded the coordinates (RA, DEC), the HD number (if available), the spectral type (SpT) and the RV value for the stars in each catalogue. 
The matching was performed using the HD identification when available or sky coordinates.
 
We also searched for the RV data of the cluster stars located in the studied area from the WEBDA database\footnote{http://www.univie.ac.at/webda/navigation.html} (and their references). We put them together with the VIZIER stars, generating a file containing all stars in the area, a total of 314 stars. For each star in this file, we have between one and six RV measurements. Due to the differences between the RV values compiled for the same star, we considered the median of all values as representative of the central RV value for each star.
In Table \ref{tab:data} we can see the collected RV data and its median value for each star, as well as our identification ID, (RA, DEC, $l$, $b$) coordinates, HD number when available and the name of the cluster hosting the star, if any.
We have also attained SpT and object types from the SIMBAD database\footnote{http://simbad.u-strasbg.fr/simbad/} for each star, where they are catalogued as: 75\% single stars, 7\% binaries, 5\% carbon stars, and the remaining 13\% are other types such as pulsating, variable, high proper motion, emission-line, blue super-giant or Be stars. 
The uncertainty of these values is highly variable and, apart from the different internal precision of each catalogue, the larger uncertainty for some RV values is due to the binary character of some objects or to the high rotational velocity of the massive stars. However, even this high uncertainty that exists in some cases does not hide the great RV dispersion inside this star-forming region, neither its kinematic pattern.

\begin{figure}
\centering
\includegraphics[width=\columnwidth]{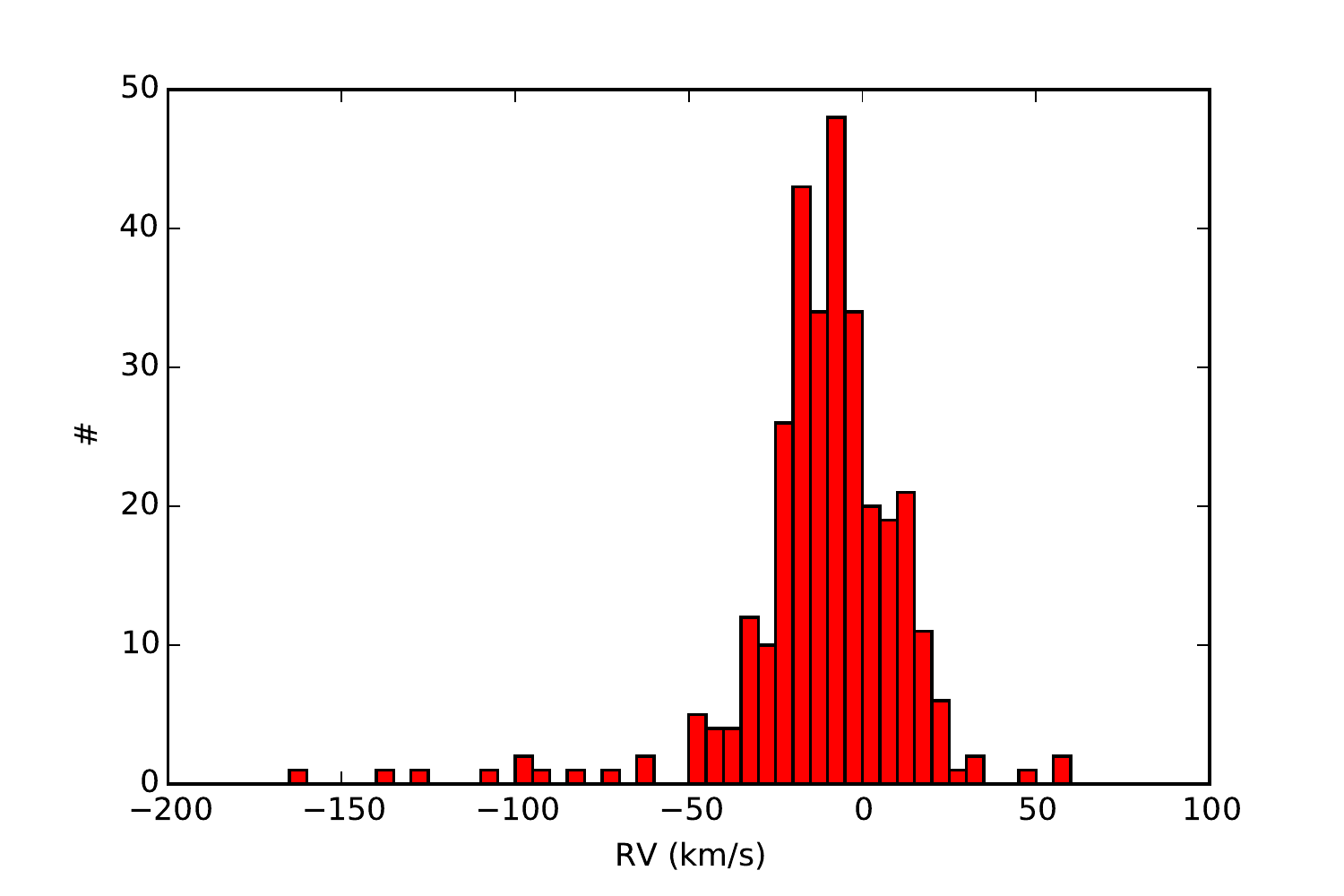}
\caption{Histogram of the RV median values calculated from the catalogues and data found in the literature. In this figure, there are two peaks at -7.5 and -17.5 km/s,  and also a plateau between 0.0 and 15.0 km/s. A negative high velocity tail is also present with -160 $<$ RV $<$ -50 km/s.}
\label{fig:histomedian}
\end{figure}

\begin{table*}
	\caption{The compiled data of each star. The columns RV + catalogue number are taken from VIZIER (references below) and the column WEBDA is the median value calculated using all measurements taken from WEBDA database (references below for each cluster). We also show our identification number, the Equatorial and Galactic coordinates, HD number, and the RV median value, which we will use in the kinematic analysis, and the name of the cluster to which the star belongs. The full table is available online.}
	\label{tab:data}
	\scalebox{0.85}[0.95]{
	\begin{tabular}{llllllllllllll} 
	\hline
 ID & RA  & DEC  & $l$  & $b$   & HD & RV\footnotemark[1]  & RV\footnotemark[2]    & RV\footnotemark[3] & RV\footnotemark[4]& RV\footnotemark[5] & WEBDA & MEDIAN & CLUSTER  \\           
& (deg) & (deg) & (deg) & (deg) & & (km/s) & (km/s)& (km/s)& (km/s)& (km/s)& (km/s)&  (km/s) \\
	\hline
 1     & 301.27500   & 38.48330   & 74.84035   & 3.65045      & 190771   & -23.60      & -24.20      & -24.20      & -999.99       & -999.99       & -999.99    & -24.20   & -  \\
 2     & 301.42500   & 35.60000      & 72.46469    & 2.00785     & 190864   & -2.80       & 3.00        & 0.00        & -21.00      & -2.00        & -15.30  & -2.40& NGC6871\footnotemark[6]       \\      
 3     & 301.50000   & 35.78330   & 72.65219    & 2.05475     & 190918   & 88.00       & -6.50       & -21.80      & -999.99       & -6.00       & -7.00   & -6.50  & NGC6871\footnotemark[6]            \\        
 4     & 301.47500   & 35.66667   & 72.54283    & 2.00932       & 190919   & -15.00      & -22.00      & -16.00      & -999.99       & -15.00      & -12.50  &  -15.00       & NGC6871\footnotemark[6]   \\           
 5     & 301.55000   & 35.38330   & 72.33638    & 1.80569     & 190967   & -16.10      & -16.10      & -999.99       & 69.00       & 49.00         & -999.99   & 16.45   & - \\
	\hline
	\end{tabular}}
\footnote[1][\cite{BA2000}; \footnote[2][\cite{AB1972}; \footnote[3][\cite{WI1953}; \footnote[4][\cite{FE1996}; \footnote[5][\cite{EV1967}; 
\footnote[6][For NGC 6871: \cite{PL1924}; \cite{CA1928}; \cite{HA1932}; \cite{PE1961}; \cite{BA1965}; \cite{CR1974}; \cite{CO1977}; \cite{SO2004}; 
\footnote[7][For NGC 6913: \cite{HA1932}; \cite{LI1989,LI1991}; \cite{GL1991}; \cite{BO2004}; 
\footnote[8][For IC 4996: \cite{HA1932}; \cite{DE1999}; 
\footnote[9][For Berkeley 86: \cite{FO1992}; \cite{HU2006}\\
\end{table*}

There are 17 objects catalogued as carbon stars in the analysed area (spectral type C or Ne). Most of them are associated with the Cygnus OB3 association and only a few of them with Cygnus OB1. 15 stars are in the general catalogue of Galactic carbon stars \citep{AL2001}. There is only one previous study concerning carbon stars in the Cygnus OB2 and OB9 associations \citep{MC1960}, but not within our current area. 

Fig. \ref{fig:histomedian} shows the histogram of the RV median values, which displays a complex structure. We can see two peaks centred at -7.5 and -17.5 , as well as a small plateau around the value of 10 km/s. 
The distribution is quite asymmetric with a low-velocity tail, which extends between -160 < RV < -50 km/s containing few stars. 
A more detailed analysis of the RV distribution is performed in section \ref{sec:structure}.

\subsection{Stellar clusters in the analysed area}
\label{sec:objects} 

\begin{figure}
\hspace{-1cm}
\includegraphics[scale=0.65]{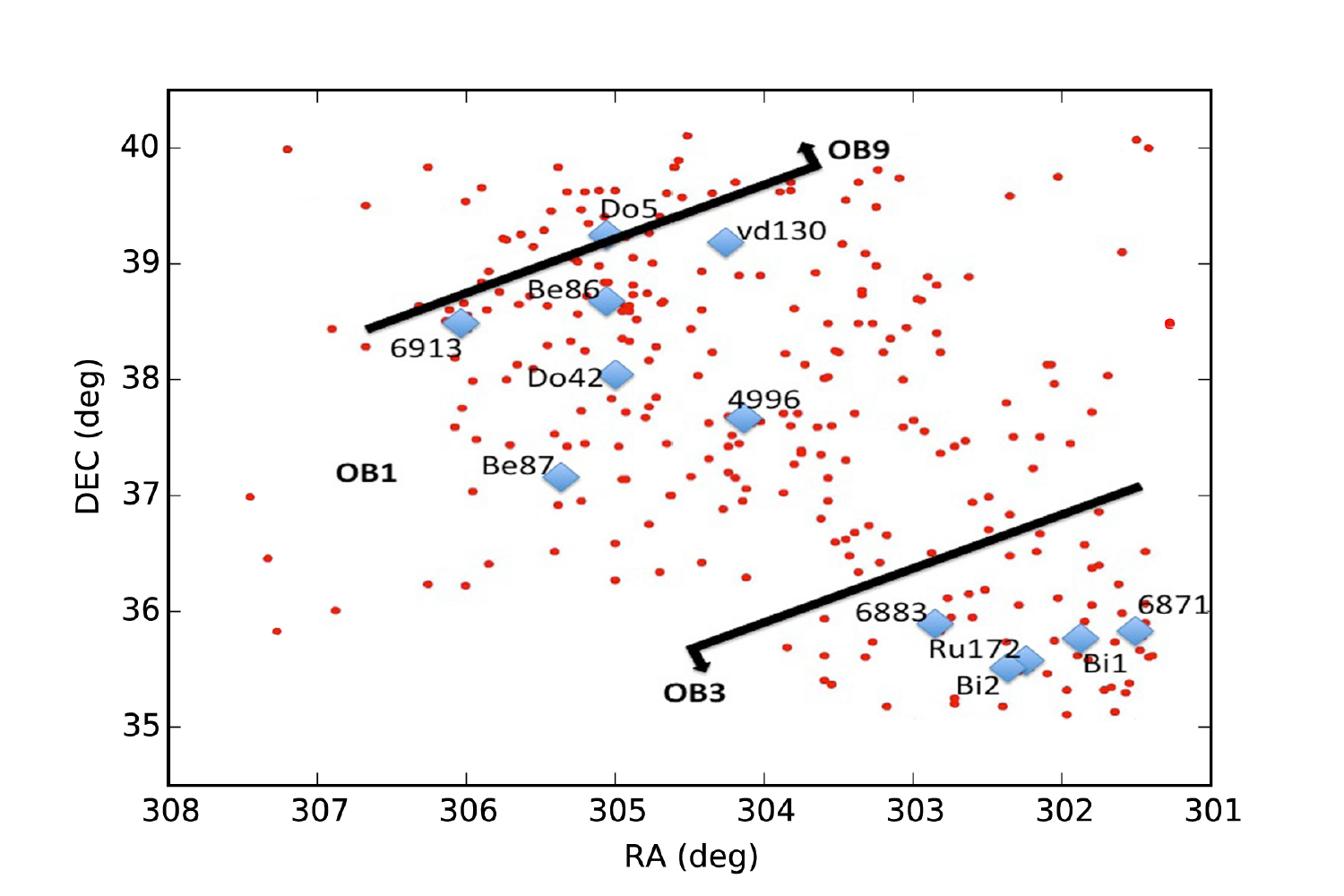}
\caption{Sky map showing the complete star sample (red dots) and the stellar clusters in the studied area (blue pentagon). The black lines mark the by-hand borders between the Cygnus OB1, OB3, and OB9 associations.}
\label{fig:radec}
\end{figure}

As mentioned above, we selected an area of 5 x 5 square degrees centred on the Cygnus OB1 association. In this field of view, 13 stellar clusters are catalogued. 
The summarized information about these objects is in Table \ref{tab:clusters}, where the coordinates, the parent OB association, the central RV values given by WEBDA and SIMBAD, respectively, and other cluster parameters like distance, reddening and age taken from four different studies \citep{LO2001, DI2002, KH2005, SI2015} are shown. The cluster positions are drawn in Fig. \ref{fig:radec} as blue pentagons. In this figure, we have also drawn approximated borders by hand between the Cygnus OB1, OB3 and OB9 associations taking into account the shape of cloud in this area and the cluster positions belonging to each association. We have six clusters inside OB1, one within OB9, and five included in the Cygnus OB3 association.

Three clusters have distances lower than that assumed for the Cygnus OB1 association (1 - 1.8 kpc). These clusters are Berkeley 87, Dolidze 5 and Dolidze 42, but the latter two are close to the lower distance limit. We found several distance estimations for the cluster Berkeley 87. \cite{LO2001} give a distance of 633 pc, while \cite{TU1982} provide a distance of 946 $\pm$ 26 pc. \cite{LO1990} conclude that the IR structure associated with Berkeley 87 is a dust shell at 950 pc, and also connected to the active region ON 2 inside the association, and that the physical connection with this association couldn't be excluded. For that reason, we didn't exclude this object from the list of clusters inside the Cygnus OB1 association. 
The cluster vdBergh 130 is located at the upper distance limit (1.8 kpc), according to \cite{SI2015}, together with IC 4996, which is also located close to the farther border of the association.

\begin{table*}
\center
	\caption{Data of the stellar clusters present in the analysed area. The parent associations, coordinates and RV1 are taken from the WEBDA database, the RV2 value is given by the SIMBAD service, and the rest of the parameters are distance, reddening and age. The specific references are shown in each case.}
	\label{tab:clusters}
	\hspace{-1cm}
	\scalebox{0.95}[1.06]{
	\begin{tabular}{lllllllllllll} 
	\hline
	Cluster & Name & Cygnus & $l$ & $b$ & RA & DEC & RV1 & RV2 &dist  & E(B-V) & Log Age\\
         & for figures & association & (deg) & (deg) & (h:m:s) & (d:m:s) &  (km/s) & (km/s)&(pc) & (mag) &   \\
	\hline
	NGC 6871 & 6871&  OB3 & 72.645 & 2.054 & 20:05:59 & +35:46:36 & -10.54\footnotemark[5] & -10.54\footnotemark[4]& 1574\footnotemark[2]  & 0.443\footnotemark[2] & 6.958\footnotemark[2]\\
	BIURAKAN 1 & Bi1& OB3 & 72.740 & 1.760 &  20:07:30 & +35:42:00 &   -7.970\footnotemark[5] & -8.480\footnotemark[4]& 1600\footnotemark[4] & 0.330\footnotemark[4] & 7.250\footnotemark[4] \\
	BIURAKAN 2 & Bi2&  OB3 & 72.751 & 1.345 & 20:09:12 & +35:29:00 &  -22.00\footnotemark[5] & -22.00\footnotemark[4]& 1106\footnotemark[2] & 0.360\footnotemark[2] & 7.011\footnotemark[2]  \\
	NGC 6883 & 6883&OB3 & 73.278 & 1.175 & 20:11:19 & +35:49:54 &   -10.33\footnotemark[5] & -10.33\footnotemark[4]& 1380\footnotemark[4] & 0.300\footnotemark[4] & 7.530\footnotemark[4]\\
	RUPRECHT 172 & Ru172& OB3 & 73.110 & 1.010 & 20:11:34 & +35:35:59 &   - & -& 1100\footnotemark[4] & 0.200\footnotemark[4] & 8.910\footnotemark[4]\\
	IC 4996 & 4996&OB1 &  75.353 & 1.306 & 20:16:30 & +37:38:00 & -18.75\footnotemark[5] & -18.75\footnotemark[4]&   1732\footnotemark[2] & 0.673\footnotemark[2] & 6.948\footnotemark[2] \\
	vdBERGH 130 & vd130&OB1 &  76.908 & 2.072 & 20:17:42 & +39:21:00 &  - & -&  1800\footnotemark[9] & 0.790\footnotemark[9] & 7.000\footnotemark[9]\\
	DOLIDZE 42 & Do42&OB1 & 76.122 & 1.065& 20:19:42 & +38:08:00 &  - & -& 972\footnotemark[2] & 0.571\footnotemark[2] & 7.542\footnotemark[2]\\
	DOLIDZE 5 & Do5&OB9 &  77.241 & 1.644& 20:20:30 & +39:23:00 &  - & -35.00\footnotemark[7]& 969\footnotemark[3] & 0.521\footnotemark[3] & 8.100\footnotemark[3]\\
	BERKELEY 86 & Be86& OB1 & 76.667 & 1.272 &   20:20:24 & +38:42:00 &  -19.30\footnotemark[1] & -25.50\footnotemark[6]& 1112\footnotemark[2] & 0.898\footnotemark[2]& 7.116\footnotemark[2]\\
	BERKELEY 87 & Be87& OB1 & 75.715 & 0.304 &20:21:42 & +37:22:00 &  - & -65.00\footnotemark[8]& 633\footnotemark[2] & 1.369\footnotemark[2] & 7.152\footnotemark[2]\\
	NGC 6913 & 6913&OB1 & 76.905 & 0.594 & 20:23:57 & +38:30:30 &   -21.05\footnotemark[5] & -21.05\footnotemark[4]&1148\footnotemark[2] & 0.744\footnotemark[2] & 7.111\footnotemark[2] \\
	\hline
	\end{tabular}}
\footnote[1][\cite{RA1999}; \footnote[2][\cite{LO2001}; \footnote[3][\cite{DI2002}; \footnote[4][\cite{KH2005}; \footnote[5][\cite{KH2007}; \footnote[6][\cite{WU2009}; \footnote[7][\cite{KH2013}; \footnote[8][\cite{DI2014}; \footnote[9][\cite{SI2015}
\end{table*}

\section{Analysis of kinematic structure}
\label{sec:structure} 

The dynamical state of a system is defined by the 6 components of its phase space, 3 velocities and 3 positions. In our case, we only have a subspace of the phase space available formed by the spatial coordinates (RA, DEC) and the radial velocity (RV).

\subsection{Velocity spectra and segregated groups}

The method proposed and tested in \cite{AL2016} to look for kinematic segregation in stellar systems is based on calculating the kinematic index ($\Lambda$) and the Spectrum of Kinematic Groupings (SKG). The kinematic index $\Lambda$ measures whether a target group (in this case, a group associated with a particular velocity channel) is more spatially concentrated than an average group with the same number of stars taken from the full sample.
 
The first step is to sort the data by RV values. We then scan the ordered RV subspace with bins of equal size $b$ and a step of size $s$, so each bin is shifted $s$ positions of the ordered variable from the previous one. In other words, ($b - s$) is the number of common objects in two consecutive bins.
For this particular case, we have chosen $s = 1$, so the coverage is the densest possible, and $b = 23$, a value close enough to the square root of the total number of objects, but increased to enhance the internal precision of $\Lambda$ without losing the information contained in the kinematic index, a balanced choice between precision and uncertainty.

We then calculate the Minimum Spanning Tree (MST) \citep{JA1930, PR1957} of the stars contained in each bin and estimate the concentration as the median edge length of the MST. We estimate the kinematic segregation index ($\tilde \Lambda$) of each bin as the ratio between the mean median length of the MST of 500 random groups $\bar{\tilde l}_{rand,500}$ and the median length of the chosen bin MST $\tilde l_{bin}$:

\begin{equation}
 \tilde{\Lambda} = \frac{\bar{\tilde l}_{rand,50}}{\tilde l_{bin}}
 \label{EqLambda}
\end{equation}

\begin{figure*}
\centering
\includegraphics[scale=0.42]{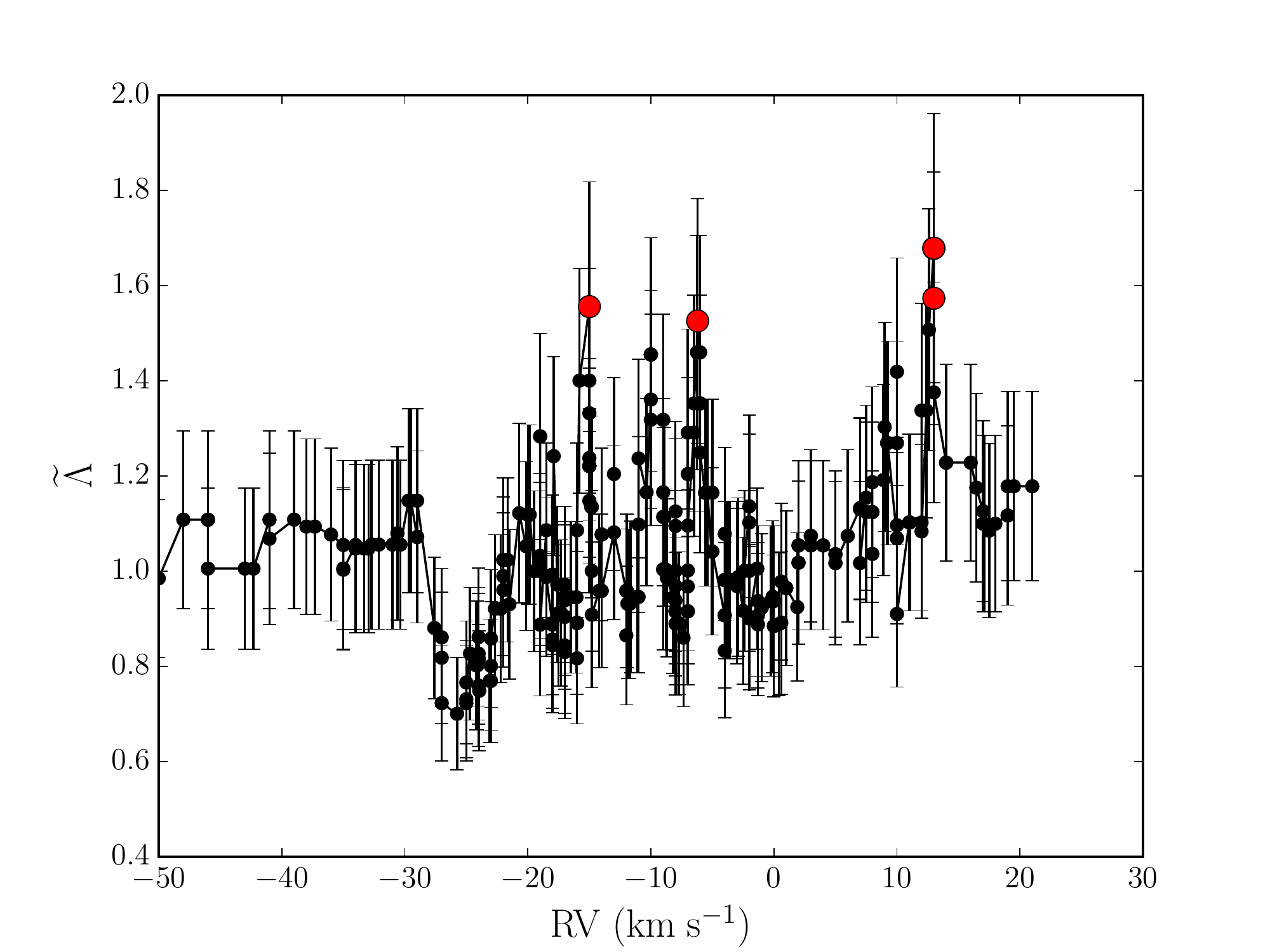}
\includegraphics[scale=0.42]{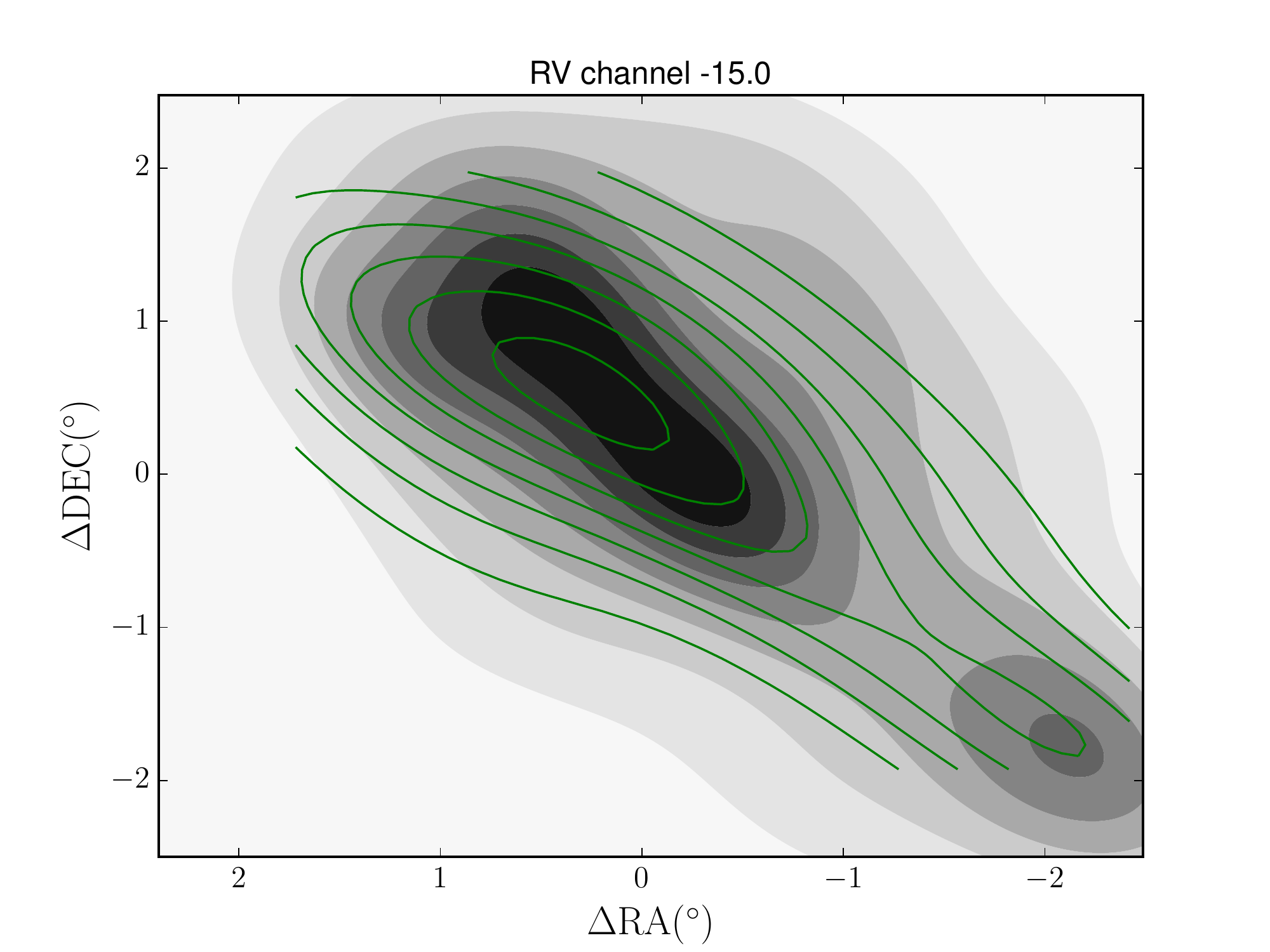}
\includegraphics[scale=0.42]{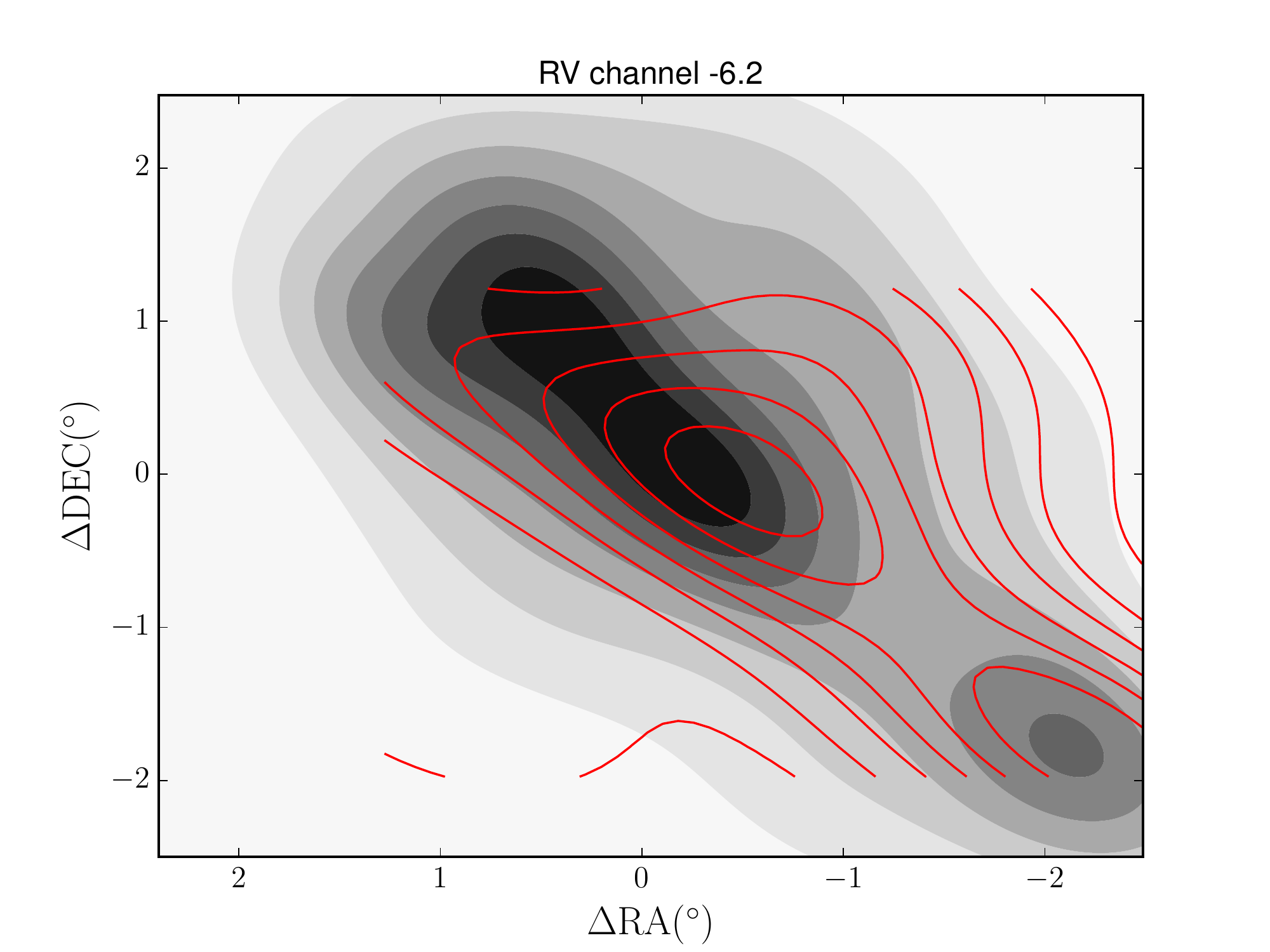}
\includegraphics[scale=0.42]{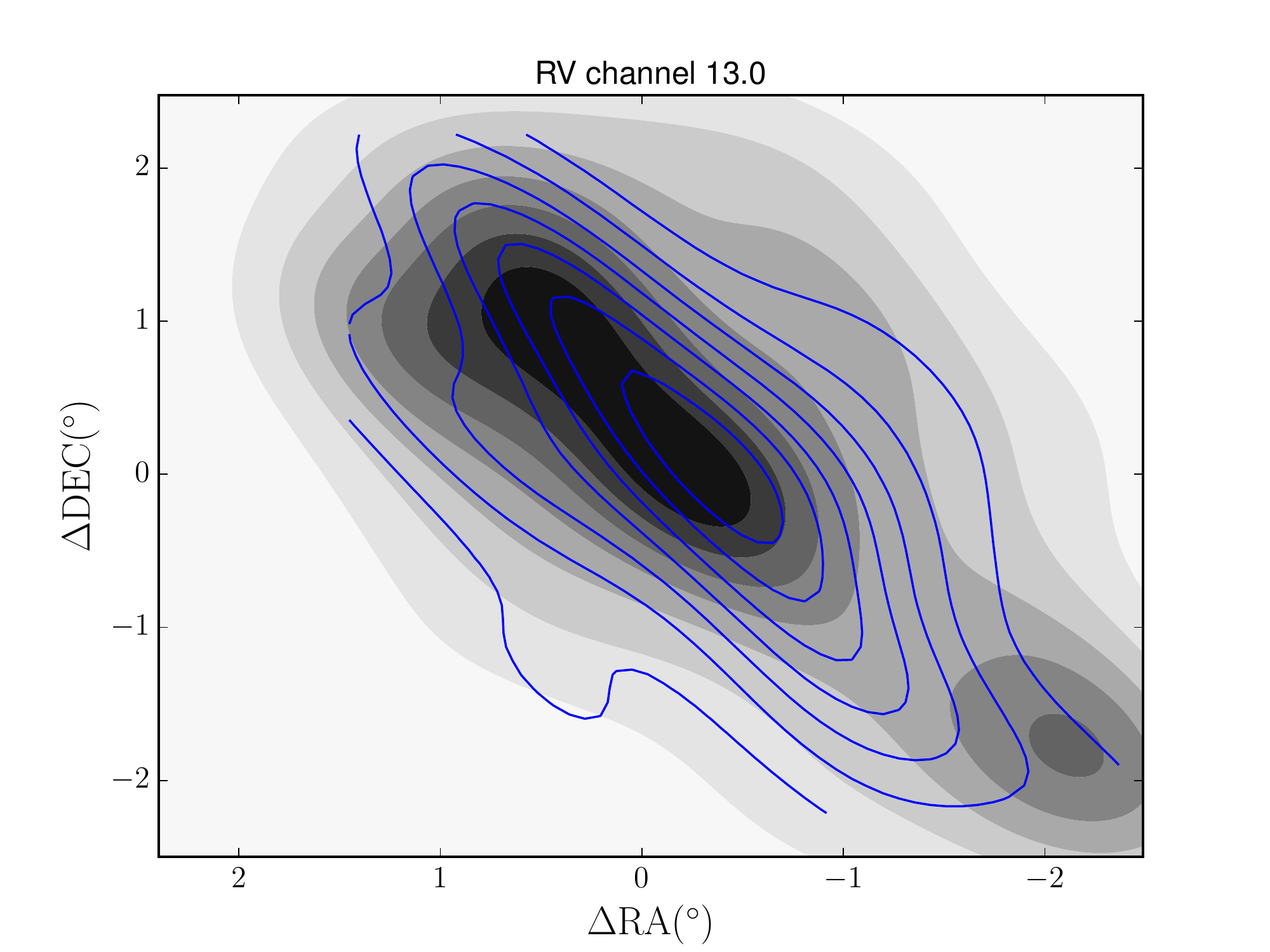}
\caption{{\it Top-left}: RV versus $\tilde{\Lambda}$ parameter showing in red the significant segregated groups around the RV median values of -15 (group 1), -6.2 (group 2) and 13 (group 3) km/s in RV. Groups 1 and 2 are formed by one bin corresponding to one red point, while group 3 has four bins, but only two red points are shown in the plot because the others are overlapping.
The next plots show the iso-density  map of the whole sample in grey scale with the iso-density contours of each segregated grouping: group 1 ({\it top-right}), group 2 ({\it bottom-left}) and group 3 ({\it bottom-right}). }
\label{fig:spectre}
\end{figure*}

When we plot $\tilde \Lambda$ against the RV median of each bin, we obtain the kinematic segregation spectrum of a cluster snapshot (see Fig. \ref{fig:spectre} {\it top-left}).

We use the standard deviation of the median edge length of 500 random groups of size $b$, $\sigma_{rand,500}$, to calculate the deviation of the $\tilde \Lambda$ value and get a conservative criterion to choose the kinematically segregated groups in the cluster. Those bins that verify:

\begin{equation}
 \frac{\bar{\tilde l}_{rand,500}}{\tilde l_{bin}} - 2 \frac{\sigma_{rand,500}}{\tilde l_{bin}} = \tilde{\Lambda} -2 \sigma_{\tilde{\Lambda}}> 1
 \label{EqCriterio}
\end{equation}

will be considered to have an index significantly above unity and thus they constitute a significant velocity segregated group. Throughout this work, we will call the bins fulfilling inequality by equation \ref{EqCriterio} segregated or selected bins.

Fig. \ref{fig:spectre} ({\it top-left}) shows the kinematic spectrum of the whole dataset calculated using the above-mentioned method with bin size $b = 23$ and step $s = 1$. Error-bars represent the deviation $\sigma_{\tilde{\Lambda}}$ of each bin as described in Eq. \ref{EqCriterio}, and red dots are the segregated bins. There are three clear groups of segregated bins, with associated RV median of -15.0 km/s (group 1), -6.2 km/s (group 2), and 13.0 km/s (group 3). There are other peaks in this kinematic spectrum, close to RV values of -20.0 km/s, -10.0 km/s and 10.0 km/s, but their $\tilde{\Lambda}$ index is not statistically significant according to our criterion in Eq. \ref{EqCriterio}. All segregated groups have a $\tilde{\Lambda}$ index larger than 1.5, meaning that segregated bins are, at least, 2/3 times more concentrated than any random group with 23 stars. 
The velocity difference between group 3 and the other significant groupings is larger than 20.0 km/s, suggesting that this stellar concentration is not part of the Cygnus OB1 association but a background or underlying population. We will come back to this point below.

We are going to analyse the three segregated groups more thoroughly. Their coloured spatial iso-density contours are shown in Fig. \ref{fig:spectre}. Each colour represents a different phase-space grouping, drawn over the spatial density distribution (grey map) of the whole sample. This grey map shows a double-peaked elongated distribution associated with the Cygnus OB1 association, with a third local density maximum well detached to the south-west of the main concentration, which corresponds to the location of the Cygnus OB3 association.
Group 1 is composed of only one bin (and thus has 23 members) and is represented in green ({\it top-right}); group 2 also has one bin and is shown in red ({\it bottom-left}); and group 3 is composed of four consecutive bins (with a total of 26 probable members) shown in blue ({\it bottom-right}). 
We wish to note that the procedure applied to detect the clustered velocity channels in the phase-space subspace does not remove the possible outliers, which could be contaminating the groups - that is, stars within the same radial velocity bin, but well separated from the spatial concentration core.   
Groups 1 and 2 are also shown as coloured spatial iso-density contours in Fig. \ref{fig:contour} - {\it top}.

\subsection{Clustering in the phase space: OPTICS}

We now consider the results of a machine learning algorithm to analyse clustering in the available 3D subspace of the phase space. We use a variation of the DBSCAN tool \citep{ES1996}, the OPTICS algorithm \citep{AN1999}, which provides a more detailed view of the clustering structure in the analysed subspace at different scales.
For an in-depth description of both methods we refer to the above cited works.

DBSCAN (Density Based Spatial Clustering of Applications with Noise) is a density-based clustering algorithm that groups points whose density is above a certain threshold. This threshold is determined by two parameters, a minimum number of neighbours, $N_N$ and a limit distance, $\epsilon$. For each cluster in the data, its core members must have at least $N_N$ neighbours within a radius $\epsilon$. The neighbours of a core member of a cluster also belong to the cluster as border points, which do not necessarily have $N_N$ neighbours in an $\epsilon$ vicinity (if they did have $N_N$ neighbours, they would be core points). DBSCAN is a well-established data mining algorithm which, amongst other advantages, does not need the user to provide a number of clusters in advance. The parameters $N_N$ and $\epsilon$ are critical as they control the required density and, although there are proposed methods for choosing $\epsilon$ \cite[see e.g.][]{ES1996, RA2016}, the suitable value is highly dependent on the specific set of data. 

The algorithm OPTICS (Ordering Points To Identify the Clustering Structure) shares the basic ideas of DBSCAN that we have just presented, but was created to detect meaningful clustering structures in data of varying density, with structures of different scales. First of all, it is important to note that it is not a clustering algorithm but a reordering of the database representing the clustering structure, but it can be used to assign points to different clusters. Given $N_N$ and $\epsilon$, OPTICS reorders the dataset where the closest points are neighbours in the new ordering, and calculates a reachability distance $\epsilon_r$, representing the density needed (once $N_N$ is fixed) for both points to be on the same cluster. The $\epsilon$ parameter, which we provide OPTICS with, can be seen as the maximum reachability distance. The results of the algorithm are represented in a reachability plot, which is a dendrogram representing the reachability distance against the OPTICS order. Denser groupings are seen as deeper valleys in the reachability plot. Even though OPTICS gives only the order and the reachability distance, we can assign the data to clusters if we consider a cutoff distance $\epsilon_c$ so adjacent points with smaller reachability distance are assigned to one cluster. 
The structures obtained with OPTICS and a cutoff distance $\epsilon_c$ are equivalent to those obtained with DBSCAN using the same number of neighbours and $\epsilon = \epsilon_c$. 
The advantage of OPTICS is the overview of clustering in different scales we get with the reachability plot, that allows a more educated choice for $\epsilon_c$.

\begin{figure}
\centering
\includegraphics[width=\columnwidth]{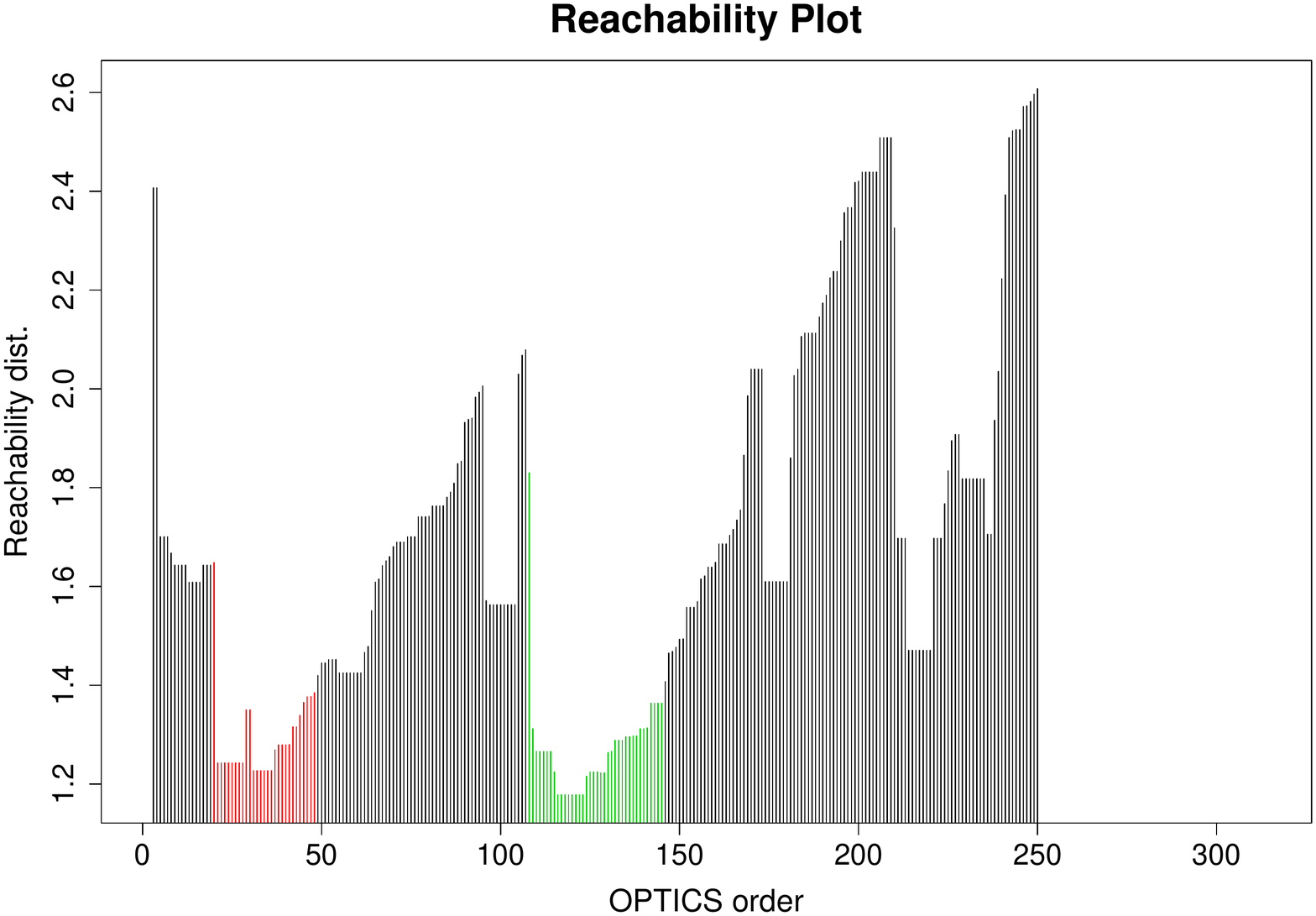}
\includegraphics[width=\columnwidth]{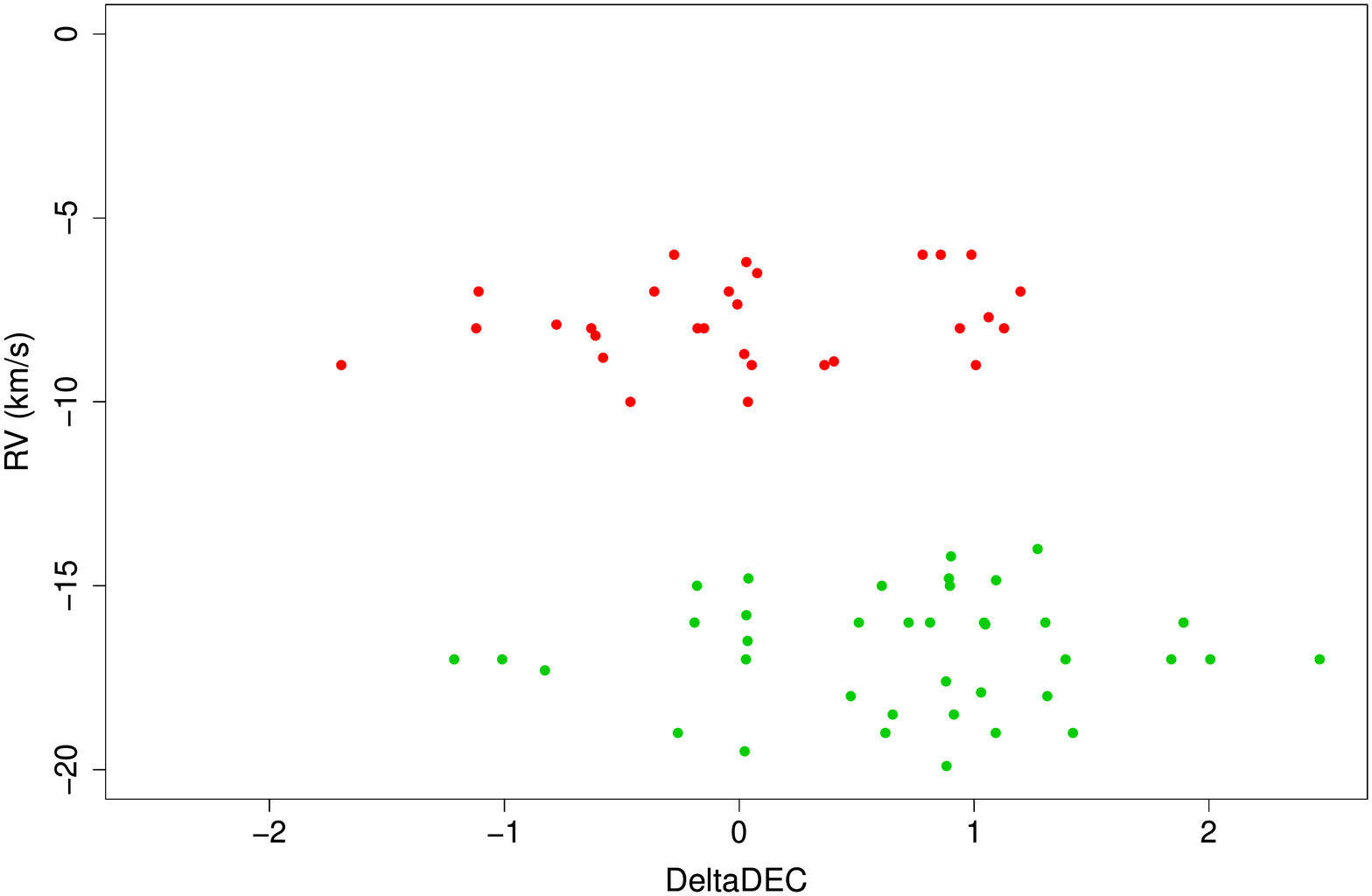}
\caption{{\it Top}: The reachability plot obtained with the OPTICS algorithm, with $N_N = 10$ and $\epsilon = 2.83$. The red and green parts of the dendrogram show the substructures obtained using a cutoff distance $\epsilon_c = 1.4$. {\it Bottom}: DEC versus RV plot for the stars corresponding to the substructures (NE - green and SW - red) for this cutoff. See the text for more details about these groupings.}
\label{fig:cutoff}
\end{figure}

In this paper, we consider the phase subspace ($\Delta$RA, $\Delta$DEC, RV), and the euclidean distance of the normalized variables. We show the results obtained using the OPTICS algorithm with $N_N = 10$ and $\epsilon = 2.83$, determined with the method proposed  by \cite{RA2016}.
High values of the cutoff distance $\epsilon_c$ lead to the detection of larger structures with a lower density, while lower $\epsilon_c$ values provide denser clusters with shorter sizes. For our particular case, we have chosen a cutoff value of 1.4, which leads to the selection of two well-defined clusters, clearly separated in the sky as well as by their central RV values. 
The coloured parts of the reachability plot (Fig. \ref{fig:cutoff} - {\it top}) show the two substructures obtained using this cutoff. 
In Fig. \ref{fig:cutoff} - {\it bottom} we can see the radial velocities associated with them.
The green substructure has a total of 38 members with RV median of -17.0 km/s, while the red has 29 members with RV median of -8.0 km/s.  

The substructures are also shown as coloured spatial iso-density contours in Fig. \ref{fig:contour} - {\it bottom}
where one of them is clearly in the north-eastern part of the Cygnus OB1 association (NE - green), and the other in the south-western part (SW - red). 
When we compare these substructures with the map in Fig. \ref{fig:radec}, which shows the stellar clusters inside the analysed area, we observe that the NE substructure has members belonging to the stellar clusters Berkeley 86 and NGC 6913, and is approximately centred on the cluster Dolidze 42, while the SW substructure has a significant number of members in IC 4996, which is close to its central position. If we consider these clusters (Dolidze 42 and IC4996) as representative of the two stellar groupings and we adopt the distance difference of the clusters as the distance between both groupings, we could say that these substructures are separated by 760 pc in the third dimension along the line of sight. 

It must be borne in mind that the distances to the Cygnus OB1 association clusters appear to show greater uncertainty than other clusters located in other Galactic regions for which the photometric parallax is more precise. The distribution of dust in the region could be the cause of this imprecision in the distance, as \cite{ST2015} have recently concluded with their reanalysis of the cluster NGC 6913 (M29).

\section{Kinematic Populations in the Cygnus OB1 association}

\subsection{Two stellar populations and two cavities}

\begin{figure}
\centering
\includegraphics[width=\columnwidth]{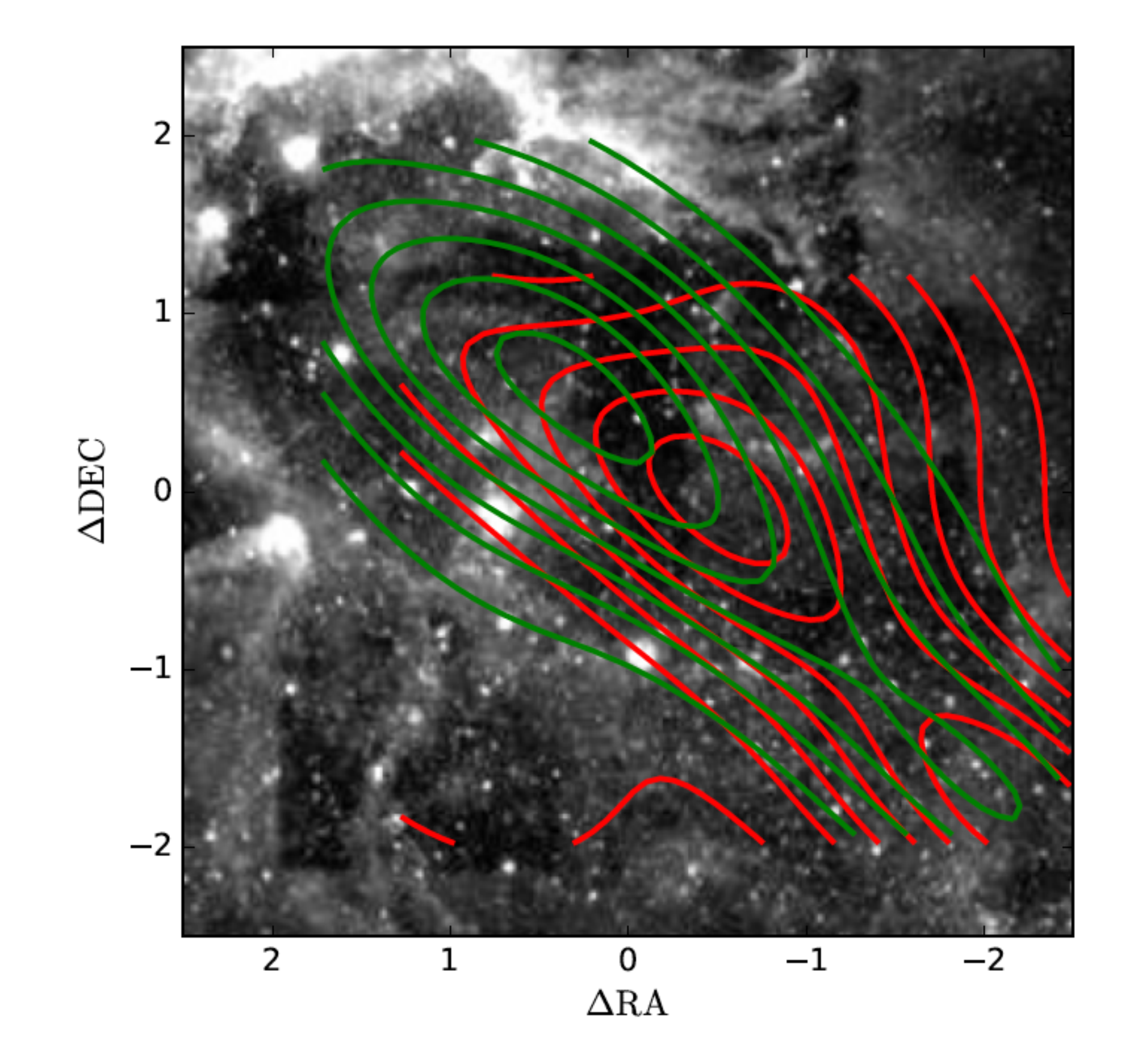}
\includegraphics[width=\columnwidth]{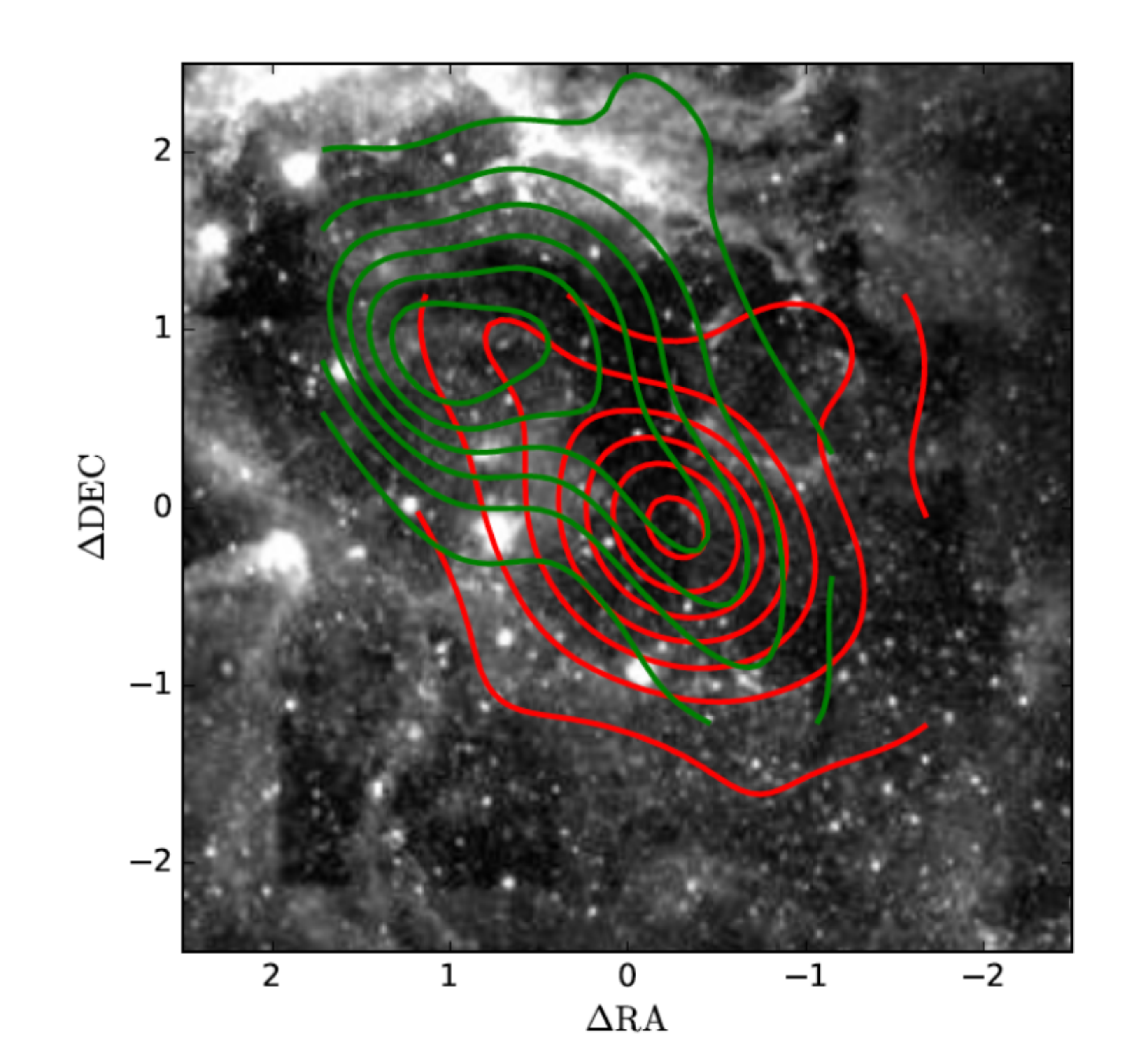}
\caption{ALLWISE black and white image with over-plotted coloured contours showing: groups 1 (green) and 2 (red) using the SKG methodology ({\it top}), the NE (green) plus SW (red) substructures corresponding to the OPTICS method({\it bottom})}
\label{fig:contour}
\end{figure}

Two different algorithms for the search for groupings in the phase subspace coincide in finding two substructures whose central values both in position and in radial velocity lead us to believe that these are the same concentrations. The SKG method shows three groupings, while OPTICS provides us with two substructures. The centroid, RV median, RV in the LSR system, the cluster closer to the centre of the grouping and its corresponding distance are shown in the Table \ref{tab:methods}. Although the mathematical algorithms utilised are based on different approaches to the problem, the results are sufficiently similar to be able to deduce that the groupings detected by the two independent methods applied basically represent the same structures. The isolines of spatial density of the two groups detected by each method are shown in Fig. \ref{fig:contour}, projected onto the ALLWISE colour image taken from the Aladin server\footnote{http://aladin.u-strasbg.fr/}, transformed into black and white \citep[][The Wide-field Infrared Survey Explorer - WISE]{WR2010}

Looking at the {\it top} panel of Fig. \ref{fig:contour} from a purely morphological point of view, the stellar population of group 1, associated with the most negative RV, seems to fill a bubble located in the north-east part of the image, with well-defined borders. Similarly, group 2, located in the south-west region of the association, appears to be more extended and to cover the rest of the area occupied by the Cygnus OB1 association. The OPTICS method and the selection of the parameter $\epsilon_c$ provide a denser and more packed version of the detected substructures (Fig. \ref{fig:contour} - {\it bottom}) than that produced by the SKG methodology, although the latter is more intuitive and easier to program and use than OPTICS.

\begin{table}
\center
	\caption{The centroid, RV median, RV in the LSR system and the cluster closer to the centre together with its distance for each grouping detected by SKG and OPTICS methods.}
	\label{tab:methods}
	\scalebox{0.7}[0.8]{
	\begin{tabular}{lllllllll} 
	\hline
	Group & Method & RA & DEC & RV & RV$_{LSR}$ &Central & Distance \\
         && (h:m:s) & (d:m:s) &  (km/s) & (km/s)&Cluster&(pc)   \\
	\hline
	1 &SKG& 20:16:42 & +38:07:59 & -15.0 & -0.3 &Dolidze42 & 972\\
	 &OPTICS& 20:19:57 & +38:30:04 & -17.0&-2.3 & & \\
	\hline
	2 &SKG& 20:14:30 & +37:33:00& -6.2 & 8.5  & IC4996 & 1732 \\
	 &OPTICS& 20:16:24 & +37:39:04& -8.0 & 6.7 &&\\
	\hline
	3 &SKG&20:16:30 &+37:39:48&13.0&27.7& &   \\
	\hline
	\end{tabular}}
\end{table}

We calculate the RV values in the Local Standard of Rest (LSR) system for our detected clustering groups using both methodologies, using the solar motion given by \cite{EL2006}. The RV difference in the LSR system for our groupings is $\sim$ 9 km/s. 
If we consider a differential rotation model as the basis of the kinematics of the Galactic disc, the distance between the groupings calculated from the difference in velocity is approximately 1.1 kpc. We wish to stress that the kinematics of a star-formation region like Cygnus OB1 are not only governed by the Galactic potential but also by the injection of energy and moment caused by the early evolution of massive stars. Therefore, this distance of separation between both kinematic groupings we have calculated may present a high uncertainty and may have been overestimated. Nonetheless, both the photometric distance that we assume (760 pc as the difference between the distances of the cluster closer to the centres of groupings), and that determined by a Galactic rotation model (1.1 kpc), undoubtedly indicate that both groups are well separated along the line of sight. Thus, the Cygnus OB1 association appears to present a NE-SW velocity gradient whose origin allows various hypotheses but whose true nature is unknown for now.

If we compare our results with those previously provided by Lozinskaya and collaborators for the ISM in the region, they probe to be compatible.
\cite{LO1988} found an extended and elongated shell and a set of smaller shells inside it, providing a hierarchical scheme of stellar formation with different forming epochs, which \cite{LO1998} shows in her figure 1, together with the massive stars' positions and the borders between associations Cygnus OB1, OB3 and OB9. The biggest cavities detected by them inside the common shell are well defined for the bright emission in $H{\alpha}$ and the radio continuum. 
Specifically, they found two kinematic components: one bright $H{\alpha}$ emission with LSR velocities in the range 2-20 km/s, and one faint $H{\alpha}$ emission with LSR velocities in two separated ranges of velocity, lower than -50 km/s and between 30 and 50 km/s, which correspond with the emission of the extended and common shell, distributed non-uniformly over that.

The two main stellar kinematic groupings detected in this study coincide morphologically with the two principal cavities detected by \cite{LO1998} in her figure 1. The stellar velocities of both groups are approximately within the velocity interval of the intense gas emission (2-20 km/s) associated with the Cygnus OB1 association, utilising the solar motion given by \cite{EL2006}, with which we calculate a mean correction to the LSR of $\sim$ 14.7 km/s. 

Finally, we want to note that the analysis using two different clustering techniques leads to the two groupings being formed by a spatial star concentration with a significant RV difference, and well separated in the three spatial dimensions. Thus, the Cygnus OB1 association contains at least two clustered stellar populations well separated in a 4D subspace of the phase space.

\subsection{Other kinematic features inside the association}

The SKG methodology detects a third grouping whose morphology looks like a filament \citep[see][for more details of the analysed test-cases]{AL2016}, which lies in the main diagonal form of the Cygnus OB1 association (Fig. \ref{fig:spectre} - {\it bottom - right}). The central RV value for this group, which we have called group 3, is of 27.7 km/s in the LSR system, a value that is close to the range observed by Lozinskaya and collaborators for the faint $H{\alpha}$ emission of the gaseous component. It could be representative of high velocity component that \cite{LO1998} called the model of two kinematic components of the shell structure. This stellar grouping could be associated with the common shell or the bubble that delimits the borders of the association. If we assume a hierarchical stellar formation model, this population could represent the oldest population of the association. 

\begin{figure}
\centering
\includegraphics[width=\columnwidth]{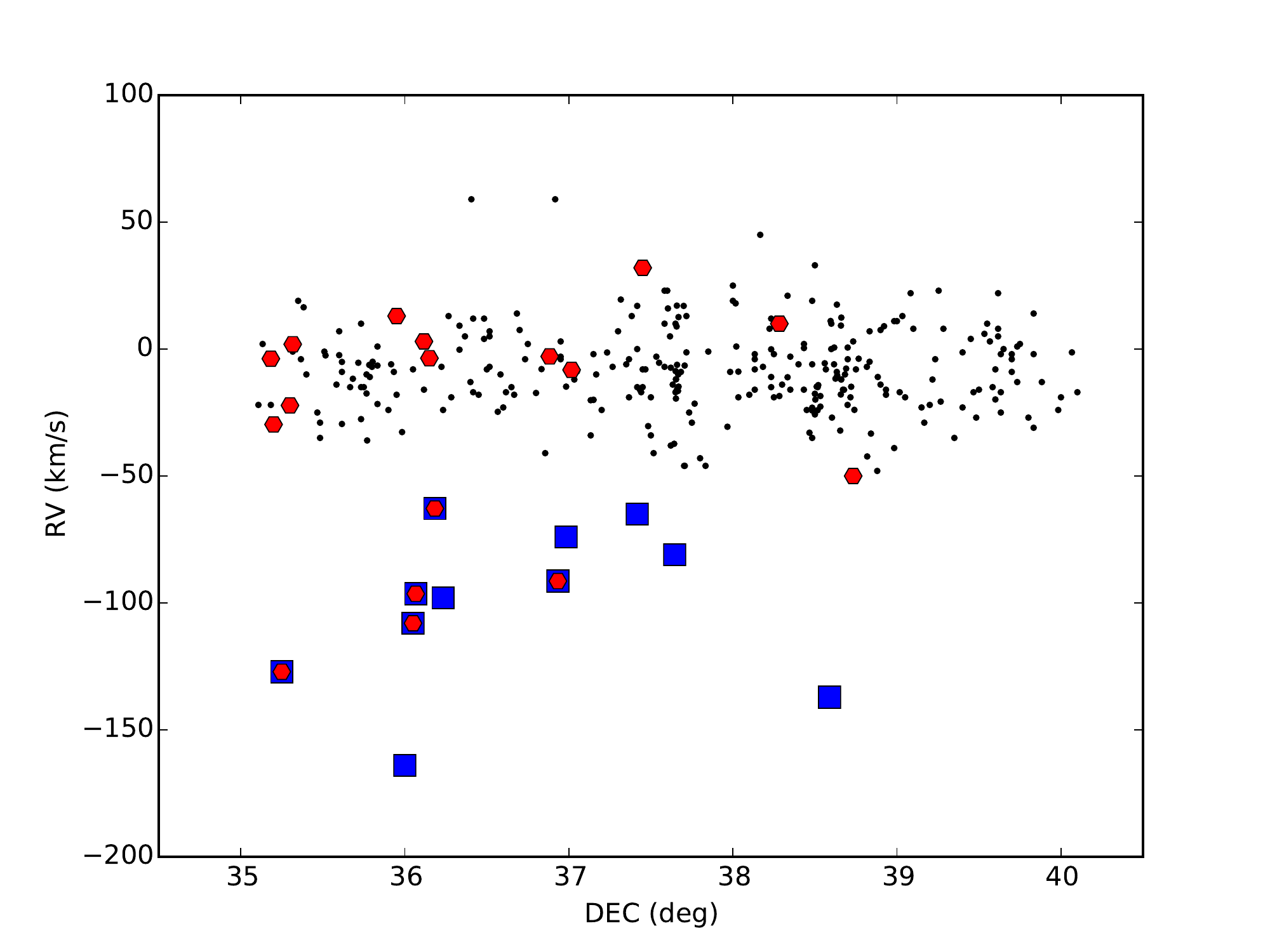}
\caption{RV versus DEC, where the black points are the whole sample, the blue squares are the stars with RV < -60 km/s and the red polygons are the carbon stars in the region.}
\label{fig:rvdec}
\end{figure}

Moreover, the RV histogram (Fig. \ref{fig:histomedian}) shows a tail of negative velocities. If we draw the RV median versus DEC (Fig. \ref{fig:rvdec}), we can see a weak but significant RV gradient with the declination for RV < -60 km/s. Most of these stars are carbon stars or have high proper motion according to the SIMBAD database and they seem to be associated with the Cygnus OB3 association. Although this apparent gradient could be an observational bias due to the complex 3D structure of the interstellar absorption in the region. The nature of this gradient, and even its very existence, is out of the scope of this paper, but we want to mention this striking observational feature.

\section{Conclusions}
The work carried out in this article can be summed up in the following points:

\begin{enumerate}[1.]

\item  Making use of the radial velocity data available in the literature for more than 300 stars within an area of 5 x 5 square degrees, we have performed an analysis of the kinematic structure of the Cygnus OB1 association. Specifically, we have focused the study on the search for groupings in the phase subspace formed by angular coordinates and radial velocity. 

\item We have applied two independent methodologies (SKG and OPTICS) to find structures in this subspace. The results obtained by both are compatible with each other, although the substructures determined through OPTICS are spatially more compact. We have found two main kinematic groupings, which have different radial velocity and different centroids, located at a significant distance from one another in the line of sight, but always within the association's assumed range of distances. We label these groupings as group 1 and group 2 according to SKG, which correspond to the NE and SW substructures according to OPTICS, respectively. The Cygnus OB1 association can be divided into two cavities, one with the walls visibles in infrared (NE part) and the other in the optical (SW part).   

\item Lozinskaya and collaborators found similar structures in their previous studies. They determined two large shells defined by ionised gas emission and another common structure that encompasses them. The elongated common shell around the Cygnus OB1 association is demarcated by bright infrared nebulae in the north-east and optical nebulae in the south-west part of the area. 

\item The third grouping (group 3) found with the SKG methodology, not detected through OPTICS, could be stars that form part of the common shell that surrounds the association and could represent the stellar population of the first starburst within the association. 

\item With regard to the distance, whether considering the photometric distance of the possible cluster centres of the two groupings, or calculating from the velocity gradient of the two detected substructures, we are led to believe that they are well separated along the line of sight. 

\end{enumerate}

Taking into account the distance determined for the two groupings, we might consider that the cavities of the Cygnus OB1 association are separated not less than a half kiloparsec along the line of sight. This distance is much larger than the size of the association over the sky, therefore the main axis of the association is along the line of sight.
How to connect this structure with the Galactic spiral structure, or even with other associations that form Cygnus-X region?. 
This region has always been controversial and the majority of studies of the spiral structure of the Galaxy \citep{GE1976, VA2011, VA2014} consider that this Galactic direction runs parallel to a spiral {\it armlet} where our sun is situated. The results found here do not contradict this hypothesis, but they do show a highly complicated field of velocities that requires more complete and precise data samples. 

The compiled data in this paper are not homogeneous, because they come from different catalogues and they were obtained using diverse instrumentation and in different epochs. We are also limited in magnitude, in that these previous studies are not particularly deep - in other words, they are unable to cover weak stars. In the near future, with the data that Gaia\footnote{http://sci.esa.int/gaia/} \citep{PE2001} and WEAVE\footnote{http://www.ing.iac.es/confluence/display/WEAV/The+WEAVE+Project} \citep{BA2010} will provide, we will be able to see whether these same structures remain or even if it is possible to find other substructures on lower spatial scales.


\section*{Acknowledgements}


This work is financed by the Spanish Ministerio de Econom\'ia y Competitividad, through grant AYA2013-40611-P.
This research has made use of the SIMBAD data base, the VizieR catalogue access tool and the Aladin sky atlas, developed at CDS, Strasbourg, France.
Also, this research has made use of the WEBDA data base, operated at the Institute for Astronomy of the University of Vienna.
This work makes use of EURO-VO software, tools or services. The EURO-VO has been funded by the European Commission through contracts RI031675 (DCA) and 011892 (VO-TECH) under the 6th Framework Program and contracts 212104 (AIDA) , 261541 (VO-ICE), 312559 (CoSADIE) under the 7th Framework Program. 
This publication makes use of data products from the Wide-field Infrared Survey Explorer, which is a joint project of the University of California, Los Angeles, and the Jet Propulsion Laboratory/California Institute of Technology, funded by the National Aeronautics and Space Administration.









\bsp	
\label{lastpage}
\end{document}